\documentclass[fleqn,usenatbib]{mnras} 

\usepackage{newtxtext,newtxmath}

\usepackage[T1]{fontenc}
\usepackage{ae,aecompl}

\usepackage{graphicx}	
\usepackage{amsmath}	
\usepackage{grffile}    

\def\Mdot{\hbox{${\dot M}$}}

\def\km{{\rm\thinspace km}}
\def\s{{\rm\thinspace s}}
\def\yr{{\rm\thinspace yr}}

\def\kmps{\hbox{${\rm\km\s^{-1}\,}$}}

\def\Msol{\hbox{${\rm\thinspace M_{\odot}}$}}

\def\Msolpyr{\hbox{${\rm\Msol\yr^{-1}\,}$}}
\def\hii{\hbox{H\thinspace{\sc ii}~}}

\title[Mass-loaded wind-blown bubbles]
{Momentum and energy injection by a wind-blown bubble into an inhomogeneous interstellar medium}

\author[J.~M.~Pittard]
{J.~M.~Pittard\thanks{E-mail: j.m.pittard@leeds.ac.uk (JMP)}\\
School of Physics and Astronomy, University of
       Leeds, Woodhouse Lane, Leeds LS2 9JT, UK
}

\date{Accepted 2022 July 07. Received 2022 July 02; in original form 2022 January 17}

\pubyear{2022}

\begin{document}
\label{firstpage}
\pagerange{\pageref{firstpage}--\pageref{lastpage}}
\maketitle

\begin{abstract}
  We investigate the effect of mass-loading from embedded clouds on
  the evolution of wind-blown bubbles. We use 1D hydrodynamical
  calculations and assume that the clouds are numerous enough that
  they can be treated in the continuous limit, and that rapid mixing
  occurs so that the injected mass quickly merges with the global
  flow. The destruction of embedded clouds adds mass into the bubble,
  increasing its density. Mass-loading increases the temperature of
  the unshocked stellar wind due to the frictional drag, and reduces
  the temperature of the hot shocked gas as the available thermal
  energy is shared between more particles. Mass-loading may increase
  or decrease the volume-averaged bubble pressure.  Mass-loaded
  bubbles are smaller, have less retained energy and lower radial
  momentum, but in all cases examined are still able to do significant
  $PdV$ work on the swept-up gas. In this latter respect, the bubbles
  more closely resemble energy-conserving bubbles than the
  momentum-conserving-like behaviour of ``quenched'' bubbles.
\end{abstract}




\section{Introduction}
\label{sec:intro}
Massive stars are key agents affecting star formation in galaxies. On
local scales they rapidly destroy star-forming molecular clouds,
through their intense radiation, powerful winds and supernova
explosions. Early (pre-supernova) feedback seems to be important,
since it is needed to explain the anti-correlation of giant molecular
clouds (GMCs) and ionized regions on 100\,pc scales and less
\citep[e.g.][]{Chevance:2022}. On larger galactic scales, supernova
feedback seems to be the dominant of the three mechanisms, determining
the amplitude of turbulent gas motions that limit and control star
formation \citep[e.g.][]{Shetty:2012}.

The importance of stellar wind feedback is still uncertain. It is
clear that the ability of a wind-blown bubble (WBB) to do $PdV$ work
on surrounding gas depends on the interior of the bubble remaining
hot. Recent work has demonstrated that turbulent mixing at the
interface between the hot interior gas and colder exterior gas can set
the cooling losses for the entire bubble. \citet{el-badry:2019} used a
1D simulation with an effective model for interface mixing and
turbulence, and found a reduction in the radial momentum of a factor
of 2. In the extreme case that the interface becomes fractal-like,
perhaps helped by perturbations due to the inhomogenity of the
surrounding gas, radiative losses might become so strong that the
bubble displays momentum-conserving-like behaviour
\citep{Lancaster:2021a,Lancaster:2021b}. Work to understand the effect
of embedded clouds on surrounding hotter gas includes analytical
studies \citep*[e.g.][]{Cowie:1977,Hartquist:1986,Fielding:2022}, and
simulations
\citep*[e.g.][]{Cowie:1981,Korolev:2015,Kim:2017,Slavin:2017,Zhang:2019,Farber:2022}.

Another issue is that numerical simulations of wind-blown bubbles have
not always had the necessary resolution for the bubble to properly
inflate. Under-resolved bubbles do not produce the correct amount of
$PdV$ work on the surrounding gas, and so have less impact on their
surroundings than they should\footnote{This issue is akin to the
  ``over-cooling'' problem that early simulations of supernova
  feedback suffered from
  \citep[e.g.][]{Katz:1992}.}. \citet*{Pittard:2021} determined the
resolution requirements for the wind injection radius to correctly
inflate the bubble.

In this work we reexamine the effect of mass-loading from embedded
clouds/clumps on the evolution of wind-blown bubbles. We note three
issues that arise in 3D simulations which directly model cloud
interactions with a larger-scale flow. First, such simulations almost
always suffer from insufficent numerical resolution, which means that
the clouds will accelerate and mix up to $5\times$ faster than they
should \citep{Pittard:2016a}. A second issue concerns cooling at
hot-cold interfaces.  \citet{Parkin:2010} showed that due to numerical
conduction, the amount of cooling is dependent on the numerical
resolution employed\footnote{In contrast, simulations of turbulent
  mixing layers by \citet{Fielding:2020} showed that the numerical
  resolution did not have a large effect on the amount of cooling. A
  definitive answer will require a correct treatment of conduction and
  the scale dependence of the fractal nature of mixing layers, which
  has not yet been fully carried out in the literature.}.  A final
issue is that when hot-cold interfaces advect across grid cells, such
as when a cold cloud surrounded by hot gas moves across the grid,
intermediate temperature grid cells are created, which then suffer
from anomously high cooling.

To avoid these issues, we do not directly model the clouds in
this work, but instead assume that they are numerous and continuously
distributed in the surrounding medium. We then assume that the clouds
that are overrun by the bubble are destroyed within the bubble
interior and inject mass into the bubble at a global rate that is
proportional to the mass-loss rate of the star\footnote{Using 3D
  numerical simulations, \citet{Rogers:2013} found that the
  mass-loading factor of a wind surrounded by a clumpy medium was of
  order several hundred.}. We assume that the flow can be treated as a
single fluid, which requires that the material liberated from the
clouds rapidly merges with the global flow and attains the same
density, velocity and temperature. 

With these assumptions, cloud destruction affects cooling in the
bubble only through the change in density and temperature associated
with the addition of (fully-mixed) mass into it, and not through
enhanced cooling at (potentially unresolved) interfaces. By minimising
cooling at hot-cold interfaces in this work, we take an opposing
position to direct simulations of WBBs in an inhomogeneous medium
(which may well overestimate the cooling). Our work follows the
same approach taken by \citet{Pittard:2019}, who investigated the
evolution of mass-loaded supernova remnants (SNRs).  In
Section~\ref{sec:setup} we note the specific details of our
calculations. In Section~\ref{sec:results} we present our results.  In
Section~\ref{sec:discussion} we discuss the validity of our
assumptions and compare our findings to previous theoretical and
observational work. In Section~\ref{sec:summary} we summarize and
conclude our work.

\section{The calculations}
\label{sec:setup}
We use a modified version of the {\sc VH-1}
code\footnote{http://wonka.physics.ncsu.edu/pub/VH-1/} to perform the
calculations. This code solves the standard inviscid equations of 1D
spherical hydrodynamics in conservative Lagrangian form, for the
conservation of mass, momentum and energy, respectively:

\begin{eqnarray}
  \label{eq:hydro_mass}
  \frac{\partial \tau}{\partial t} - \frac{\partial (r^{2}u)}{\partial
  m} = \dot{\tau}, \\ 
  \label{eq:hydro_mtm}
  \frac{\partial u}{\partial t} + r^{2}\frac{\partial P}{\partial
  m} = 0, \\
  \label{eq:hydro_energy}
  \frac{\partial E}{\partial t} + \frac{\partial (r^{2}uP)}{\partial
  m} = \dot{E},
\end{eqnarray}  
\noindent where $\tau$ is the specific volume ($\rho=1/\tau$ is the
fluid mass density), $u$ is the velocity, and $P$ is the pressure.
$E = \rho u^{2}/2 + P/(\gamma-1)$ is the total energy per unit volume,
where $\gamma=5/3$ is the ratio of specific heats. The mass coordinate
$m$ is defined by $dm = \rho r^{2} dr$, where $r$ is the radial
coordinate.

Piecewise parabolic spatial reconstruction is applied to the fluid
variables to obtain values at each cell interface. These are input
into the iterative, approximate two-shock Riemann solver of
\citet{Colella:1984}. This outputs the time-averaged fluxes at each
interface to update the fluid variables. Finally, the updated
quantities are remapped to the original grid at the end of every
step. This approach is known as PPMLR: Piecewise Parabolic Method with
Lagrangian Remap.

Two source terms are added to the hydrodynamic equations which are treated via
operator splitting. Firstly, the rate of
change of the internal energy per unit volume due to heating and
cooling is:
\begin{equation}
\label{eq:eintdot}
\dot{E} = (\rho/m_{\rm H}) \Gamma - (\rho/m_{\rm H})^{2}\Lambda(T),
\end{equation} 
where the temperature-independent heating coefficient
$\Gamma=2\times10^{-26}\,{\rm erg\,s^{-1}}$.  The cooling curve,
$\Lambda(T)$, is calculated assuming collisional ionization
equilibrium and is constructed from 3 separate parts \citep*[for
further details
see][]{Wareing:2017a,Wareing:2017b,Kupilas:2021,Pittard:2022}. We
assume solar abundances with mass fractions $X_{\rm H}=0.7381$,
$X_{\rm He}=0.2485$, and $X_{\rm Z}=0.0134$
\citep[cf.][]{Grevesse:2010}. We also use a temperature-dependent
average particle mass, which is determined from a look-up table of
values of $P/\rho$ \citep{Sutherland:2010}.

The second source term is the rate of change of the specific volume
due to cloud destruction, $\dot{\tau} = -\dot{\rho}/\rho^{2}$. We
assume that the clumps inject mass at a global rate inside the bubble
of $\dot{M}_{\rm cl} = f_{\rm ML}\,\dot{M}_{\rm w}$, where
$\dot{M}_{\rm w}$ is the mass-loss rate of the star and $f_{\rm ML}$
is the mass-loading factor which sets the relative strength of the
mass-loading. We assume that the mass injection occurs uniformly
within the bubble\footnote{Note, however, that this assumption is
  subject to the presence of available clump material, and in cases
  where the clump mass ``runs out'', the mass injection occurs only in
  the part of the bubble where clumps remain - in this scenario the
  model mimics clouds being quickly destroyed and mass-loading
  occuring only near the bubble edge.}, so that the rate of change of
the gas density is given by
\begin{equation}
  \label{eq:rhodot}
\dot{\rho} = f_{\rm ML}\,\dot{M}_{\rm w}/V_{\rm bub},  
\end{equation}
where $V_{\rm bub} = \frac{4}{3}\pi r_{\rm FS}^{3}$ is the volume of
the WBB and $r_{\rm FS}$ is the radius of the forward
shock. Mass-loading occurs inside the WBB (including in the swept-up
shell) but outside the injection region of the wind. As the bubble
expands and its volume increases, the mass injection rate per unit
volume decreases. The injected mass is assumed to be at rest and cold
before mixing with the flow so there are no momentum or energy source
terms due to the mass-loading.

The mass in the clumps has no effect on the dynamics of the bubble
other than to add mass to the bubble interior. The clouds have no
momentum imparted to them by the bubble: they are immoveable, swept-up
objects that can do nothing but evaporate/mix. In reality, clouds will
pick up momentum from the flow and move downstream. The exact distance
that clouds travel downstream before fully mixing with the flow is not
known, but will depend on such things as the cloud size and density,
the density, velocity and Mach number of the flow, whether the cloud
is smooth or structured, whether the cloud is impacted by a wind or a
shock, and whether magnetic fields, thermal conduction and radiative
cooling are important \citep*[see,
e.g.,][]{Klein:1994,Nakamura:2006,Scannapieco:2015,McCourt:2015,Bruggen:2016,Pittard:2016a,Pittard:2016b,Goldsmith:2016,Goldsmith:2017,Schneider:2017,Goldsmith:2018,Banda-Barragan:2019,Goldsmith:2020}.
In addition, the presence of other clouds can affect the interaction
\citep*[e.g.][]{Poludnenko:2002,Aluzas:2012,Aluzas:2014,Forbes:2019,Banda-Barragan:2020,Banda-Barragan:2021}. Such
complications are ignored in the current work.

An additional parameter in our calculations is the ratio of mass in
clumps to the mass in interclump gas in the background, which we
define as $\nu$. The interclump background density that the bubble is
expanding into is $\rho_{\rm ic}$. The large-scale, smoothed-out
density of clumps $\rho_{\rm cl,avg} = \nu \rho_{\rm ic}$, though by
definition the clouds have actual densities
$\rho > \rho_{\rm cl,avg}$. Larger values of $\nu$ mean that there is
a larger reservoir of cloud mass that can be injected into the
WBB. Small and/or low density clouds that are relatively rare would be
consistent with a small value of $\nu$ and high value of $f_{\rm ML}$
(rapid cloud destruction, with most of the mass injection occuring
close to the forward shock). Large and/or high density clouds are
consistent with a large value of $\nu$ (long-lived clouds, with mass
injection throughout the bubble).  We keep track of the cloud mass in
each grid cell which declines as mass is injected into the bubble.  In
regions where the local clumps are completely destroyed, which can
occur when there is rapid mass-loading and a finite reservoir of mass
in the clumps, no further clump mass is added to the grid cells
affected. In such circumstances the global rate of mass injection into
the bubble can fall below $f_{\rm ML}\,\dot{M}_{\rm w}$. We include
advected scalars to track the mass fractions of wind, injected and
ambient material in each grid cell.

We assume that the star is a single O-star, and has a mass-loss rate
$\dot{M}_{\rm w}=10^{-7}\,\Msolpyr$, a stellar wind speed
$v_{\rm w} = 2000\kmps$, and a main-sequence lifetime
$t_{\rm MS}=5\,$Myr \citep[such parameters are typical of a
``late''ish O-star, e.g.,][]{Marcolino:2022}. During this period the
star injects $0.5\,\Msol$ of mass, $10^{3}\,\Msol\kmps$ of momentum,
and $2\times10^{49}$\,erg of energy into its surroundings.

\citet{Pittard:2021} examined the effects of different wind injection
mechanisms and numerical resolution on the ability of a wind-blown bubble to
inflate. They found that the radius of the wind injection region, $r_{\rm inj}$,
needed to be significantly less than
\begin{equation}
r_{\rm inj,max} = \left(\frac{\Mdot_{\rm w} \,v_{\rm w}}{4 \pi P_{\rm amb}}\right)^{1/2},
\end{equation}
where $P_{\rm amb}$ is the pressure of the ambient medium. In this
work we adopt $r_{\rm inj}=0.01\,r_{\rm inj,max}$, which gives an
accurate value for the momentum of the bubble \citep{Pittard:2021}. We
use 5 cells for the injection radius. All simulations use the {\em
  meo} wind injection method \citep[see][]{Pittard:2021}.

We run a number of simulations to investigate the effect of varying
the values of $f_{\rm ML}$ and $\nu$ on the evolution of the
bubble. We also explore both low and high intercloud densities. Our
models are noted in Tables~\ref{tab:ML_results}
and~\ref{tab:ML_results_ramb2e-21} where we also record various
properties of the bubble at the end of the simulation, including its
radius, $r_{\rm FS}$, the swept-up intercloud mass, $M_{\rm sw}$, the
mass injected into the bubble from the clumps, $M_{\rm inj}$, the mass
of hot gas (defined as gas with $T > 2\times10^{4}$\,K),
$M_{\rm hot}$, the thermal energy, $E_{\rm th}$, the kinetic energy,
$E_{\rm kin}$, and the radial momentum, $p_{\rm bub}$. Of course, the
results of our study may be somewhat specific to the parameter values
adopted.

\section{Results}
\label{sec:results} 

\subsection{``Low'' intercloud density}
In the following calculations we adopt an interclump number density of
Hydrogen nuclei $n_{\rm H,ic} = 1\,{\rm cm^{-3}}$. This gives
$\rho_{\rm ic} = 2.267\times10^{-24}\,{\rm g\,cm^{-3}}$, a mean
molecular weight $\mu_{\rm H,ic}=1.07$, and $T_{\rm ic} =
2950$\,K. The latter two values arise from assuming thermal
equilibrium with the adopted cooling curve.  The pressure of the
intercloud gas, $P_{\rm ic} = 5.17\times10^{-13}\,{\rm dyn \,cm^{-2}}$
(or $P_{\rm ic}/k = 3750\,{\rm K \,cm^{-3}}$).  With these parameters
$r_{\rm inj,max} = 4.51$\,pc ($P_{\rm amb}=P_{\rm ic}$ in our current
work). The width of each grid cell is $dr=9.028\times10^{-3}$\,pc.

\begin{table*}
\centering
\caption[]{Physical quantities at $t=5$\,Myr in models of WBBs
  expanding into an inhomogeneous ambient medium of intercloud
  hydrogen nucleon number density $n_{\rm H,ic}=1\,{\rm cm^{-3}}$
  ($\rho_{\rm ic} = n_{\rm H,ic}m_{\rm H}/X_{\rm H} =
  2.267\times10^{-24}\,{\rm g\,cm^{-3}}$), temperature
  $T_{\rm ic} = 2950$\,K, and pressure
  $P_{\rm ic} = 5.17\times10^{-13}\,{\rm dyn \,cm^{-2}}$
  ($P_{\rm ic}/k = 3750\,{\rm K \,cm^{-3}}$). The ratio of initial
  cloud mass to intercloud mass is $\nu$. The injection of cloud gas
  into the bubble occurs at a rate
  $\dot{M}_{\rm cl} = f_{\rm ML}\,\dot{M}_{\rm w}$. The measured
  quantities are the radius of the bubble ($r_{\rm FS}$), the swept-up
  intercloud mass ($M_{\rm sw}$), the injected mass from clump
  destruction ($M_{\rm inj}$), the mass of ``hot'' gas
  ($M_{\rm hot}$), the total thermal ($E_{\rm th}$) and kinetic
  ($E_{\rm kin}$) energies, and the radial momentum ($p_{\rm
    bub}$). $M_{\rm hot}$ can sometimes show significant variation on
  short timescales: when this occurs the quantity is recorded in
  italics. Values in square brackets are normalised to the standard
  \citet{Weaver:1977} bubble assuming an adiabatic interior and a
  negligible external presssure. $E_{\rm th}$ and $E_{\rm kin}$
  measure the energy in the entire bubble (wind, swept-up and injected
  mass) but are normalized by just the thermal energy of the hot gas
  and the kinetic energy of the swept-up shell, respectively.}
\label{tab:ML_results}
\begin{tabular}{lccccccccc}
\hline
Model & $f_{\rm ML}$ & $\nu$ & $r_{\rm FS}$ & $M_{\rm sw}/10^{4}$ & $M_{\rm inj}$ & $M_{\rm hot}$ & $E_{\rm th}/10^{49}$ & $E_{\rm kin}/10^{48}$ & $p_{\rm bub}/10^{4}$ \\
      &       &             & (pc)        & ($\Msol$)   & ($\Msol$)    &($\Msol$)     & (erg) & (erg) & ($\Msol\,{\rm km\,s^{-1}}$)\\
\hline
fML0           & 0.0 & - & 51.5 [1.12] & 1.90 [1.41] & 0.0 & {\it 0.54} [1.08] & 1.62 [1.79] & 1.34 [0.35]& 4.18 [0.58]\\
fML10\_nu1e10  & 10  & $10^{10}$ & 51.5 [1.12] & 1.90 [1.41] & 5.0 & {\it 3.1} [6.2] & 1.62 [1.79]& 1.34 [0.35]& 4.18 [0.58]\\
fML100\_nu1e10  & $10^{2}$ & $10^{10}$ & 50.5 [1.10] & 1.79 [1.33]& 50.0 & 17.6 [35.2]& 1.49 [1.65]& 1.16 [0.30]& 3.75 [0.52]\\
fML1000\_nu1e10 & $10^{3}$ & $10^{10}$ & 44.0 [0.96]& 1.18 [0.88]& 499  & 20.8 [41.6]& 0.83 [0.92]& 0.23 [0.06]& 1.08 [0.15]\\    
fML10\_nu10 & 10 & 10 & 51.5 [1.12]& 1.90 [1.41]& 5.0 & {\it 3.1} [6.2]& 1.62 [1.79]& 1.34 [0.35]& 4.18 [0.58]\\  
  fML100\_nu10 & $10^{2}$ & 10 & 50.5 [1.10] & 1.79 [1.33]& 50.0 & 17.3 [34.6]& 1.49 [1.65]& 1.16 [0.30]& 3.75 [0.52]\\    
fML1000\_nu10 & $10^{3}$ & 10 & 45.0 [0.98]& 1.27 [0.94]& 498 & 21.3 [42.6]& 0.89 [0.98]& 0.23 [0.06]& 1.08 [0.15]\\    
fML1000\_nu1 & $10^{3}$ & 1 & 45.4 [0.99]& 1.30 [0.96]& 486 & 21.3 [42.6]& 0.91 [1.01]& 0.22 [0.06]& 1.09 [0.15]\\
fML1000\_nu0.1 & $10^{3}$ & 0.1 & 48.6 [1.06]& 1.60 [1.19]& 395 & 22.4 [44.8]& 1.09 [1.21]& 0.25 [0.06]& 1.40 [0.19]\\
fML1000\_nu0.01 & $10^{3}$ & 0.01 & 51.4 [1.12]& 1.89 [1.40]& 140 & {\it 0.50} [1.0]& 1.61 [1.78]& 1.35 [0.35]& 4.19 [0.58]\\
\hline  
\end{tabular}
\end{table*}

\subsubsection{No mass-loading}
\begin{figure*}
\includegraphics[width=16.0cm]{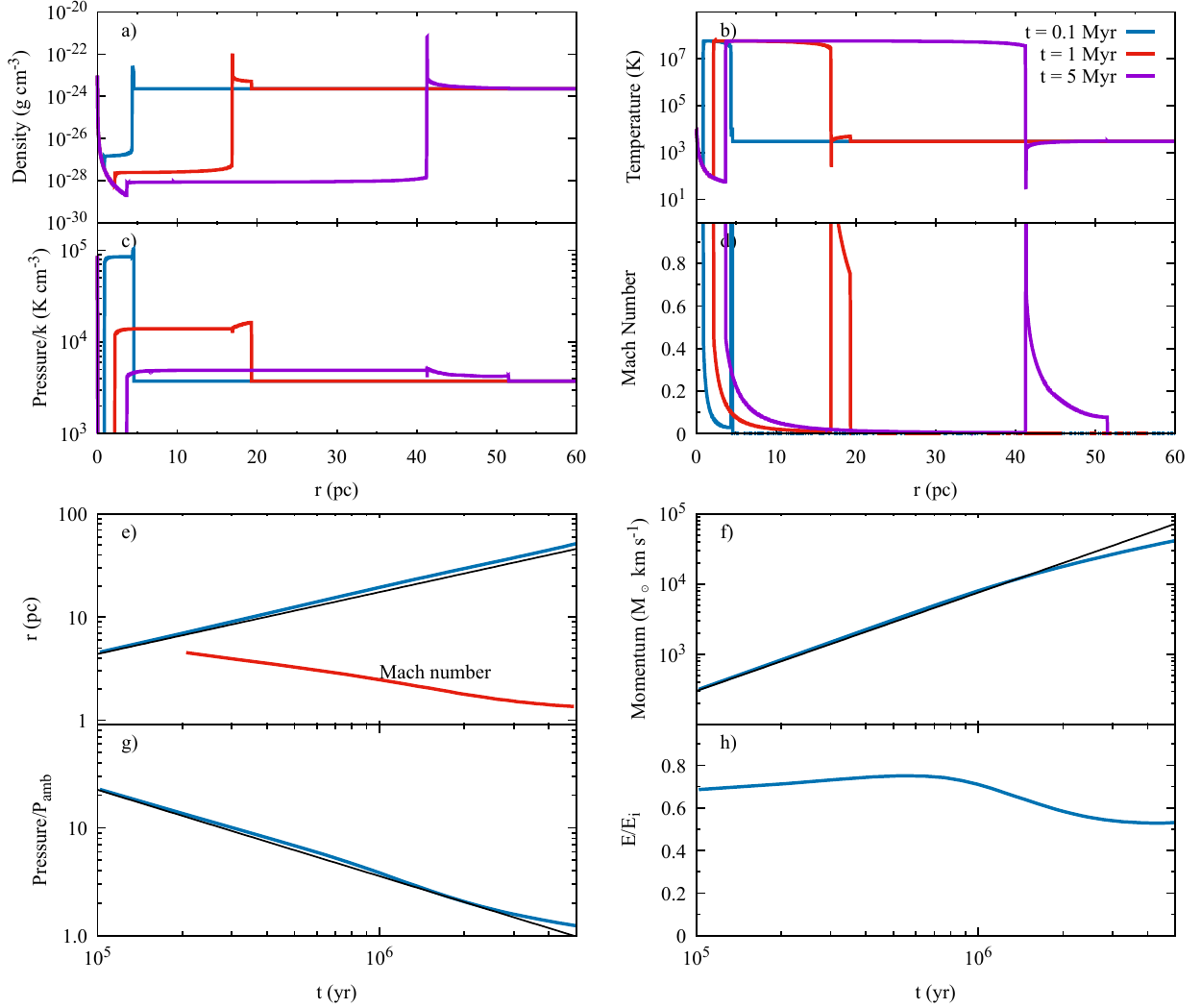}
\caption{The evolution of a bubble with
  $n_{\rm H,ic}=1\,{\rm cm^{-3}}$
  ($\rho_{\rm ic}=2.267\times10^{-24}\,{\rm g\,cm^{-3}}$) and no
  mass-loading ($f_{\rm ML}=0$). Panels a)-d) show profiles of the
  density, temperature, pressure and adiabatic Mach number at $t=0.1$,
  1 and 5\,Myr. Panel e) shows the bubble radius (and the isothermal
  Mach number of the forward shock); f) shows the radial momentum; g)
  shows the volume averaged presssure; and h) shows the retained
  energy fraction. In panels e)-h), simulation results are shown in
  blue and red, and analytical results in black.}
\label{fig:wbb_fML0}
\end{figure*}

We begin by examining the evolution of the WBB without any
mass-loading (i.e. $f_{\rm ML}=0$). Fig.~\ref{fig:wbb_fML0}a-d) shows
density, temperature, pressure, and adiabatic Mach number profiles at
3 bubble ages.  We see the classic bubble structure which consists of
freely outflowing stellar wind, a reverse shock, a region of shocked
stellar wind, a contact discontinuity, a region of swept-up ambient
material, and a forward shock. The shell formation time is
$\approx 2100$\,yr \citep[cf. Eq.~4.3 in][]{Koo:1992}. The shell is
initially very thin, due to the high Mach number of the forward shock,
but thickens as the bubble expansion slows down. In the following,
when we refer to the ``bubble'', we mean the entire entity (shocked
stellar wind, swept-up material, and in the case of mass-loading also
the injected mass).

At $t=5$\,Myr, the radii of the reverse shock, contact discontinuity
and forward shock are at approximately 3.7, 41.3 and 51.5\,pc,
respectively. The shocked stellar wind is heated to
$\approx5\times10^{7}$\,K, and its high pressure drives the expansion
of the bubble. At $t=5$\,Myr, some of the swept-up material is in a
dense and cold shell near the contact discontinuity, where the gas is
compressed to about $300\times$ the ambient intercloud density, and
cools to below 30\,K. However, due to the low isothermal Mach number
of the forward shock at this time ($\approx 1.3$), material which has
been more recently shocked is compressed very little, and a much
thicker shocked region separates the forward shock from the densest
part of the shell (at $t=5$\,Myr the thick shell is most clearly seen
in the Mach number plot in Fig.~\ref{fig:wbb_fML0}d).

The pressure within the bubble is almost constant at any given moment
in time, but drops markedly as the bubble expands. The bubble is
over-pressured by factors of approximately 20 and 4 at $t=0.1$ and
1\,Myr, respectively. The supersonic wind has a very high Mach number,
but the Mach number just after the reverse shock drops to
approximately 0.4, and declines further with radius, reaching its
lowest value at the contact discontinuity. The cooling of the swept-up
gas means that the adiabatic Mach number increases between the forward
shock and the contact discontinuity.

Fig.~\ref{fig:wbb_fML0}e)-h) shows the evolution of the radius of the
WBB, its radial momentum, the pressure inside the bubble, and the
retained energy. The thin solid black line in the panels shows the
analytical solution for a bubble with a hot, adiabatic interior
(Eqs.~21 and 22 in \citet[][]{Weaver:1977} for the bubble radius and
pressure; and Eq.~9 in \citet[][]{Pittard:2021} for the bubble radial
momentum).  We see that the bubble behaves generally as expected,
though there are some slight disagreements with the analytical
theory. The differences arise because the analytical theory assumes
that the ambient pressure is much smaller than the bubble
pressure. However, we see in Fig.~\ref{fig:wbb_fML0}g) that this is
not true at late times, which is when the differences between the
theoretical and model results are at their greatest. The result is
that the forward shock transitions from an initially strong shock to a
weak shock.  Fig.~\ref{fig:wbb_fML0}e) shows that the forward shock
isothermal Mach number is 4.5 at $t=0.2$\,Myr, 2.5 at $t=1$\,Myr, and
just 1.35 at $t=5$\,Myr. This causes the bubble radius to be greater
than expected as the swept-up shell grows in relative thickness with
time. It also causes the bubble radial momentum to drop away from the
analytical value at late times (Fig.~\ref{fig:wbb_fML0}f)\footnote{We
  have confirmed that the disagreement with theory and simulation in
  Fig.~\ref{fig:wbb_fML0}e)-g) is due to the ambient pressure becoming
  significant by repeating the calculation with a lower ambient
  pressure. To achive this we artificially lowered the heating rate of
  gas with $T<6000$\,K (specifically, we multiply $\Gamma$ by a factor
  $[1- (6000-T)/6000]$ for temperatures $T < 6000$\,K). This led to a
  much reduced temperature for the ambient intercloud material
  ($T_{\rm ic}=30.4$\,K), and a comensurate drop in the ambient
  pressure ($P_{\rm amb}=2.45\times10^{-15}\,{\rm dyn\, cm^{-2}}$),
  but it does not affect the strength of cooling in the model. With
  this change we find that the bubble momentum perfectly tracks the
  analytical theory. Analytical theory for cases where the ambient
  pressure is significant can be found in \citet{GarciaSegura:1996}.}.

\begin{table*}
\centering
\caption[]{As Table~\ref{tab:ML_results} but for an inhomogeneous
  ambient medium of intercloud hydrogen nucleon number density
  $n_{\rm H,ic}=884\,{\rm cm^{-3}}$
  ($\rho_{\rm ic} = 2\times10^{-21}\,{\rm g\,cm^{-3}}$), temperature
  $T_{\rm ic}=21.2$\,K, and pressure
  $P_{\rm ic} = 1.48\times10^{-12}\,{\rm dyn \,cm^{-2}}$
($P_{\rm ic}/k = 1.075\times10^{4}\,{\rm K \,cm^{-3}}$).}
\label{tab:ML_results_ramb2e-21}
\begin{tabular}{lccccccccc}
\hline
Model & $f_{\rm ML}$ & $\nu$ & $r_{\rm FS}$ & $M_{\rm sw}/10^{5}$ & $M_{\rm inj}$ & $M_{\rm hot}$ & $E_{\rm th}/10^{48}$ & $E_{\rm kin}/10^{48}$ & $p_{\rm bub}/10^{5}$ \\
      &       &             & (pc)        & ($\Msol$)   & ($\Msol$)    &($\Msol$)     & (erg) & (erg) & ($\Msol\,{\rm km\,s^{-1}}$)\\
\hline
fML0          & 0.0 & - & 11.5 [0.97]& 1.88 [0.93]& 0.0 & {\it 0.65} [1.3]& 8.31 [0.92]& 3.35 [0.87]& 2.51 [0.89]\\
fML10\_nu1e10  & 10 & $10^{10}$ & 11.4 [0.97]& 1.85 [0.91]& 5.0 & {\it 3.7} [7.4]& 8.01 [0.89]& 3.23 [0.83]& 2.45 [0.87]\\
fML100\_nu1e10  & $10^{2}$ & $10^{10}$ & 8.51 [0.72]& 0.76 [0.37]& 49.9 & 4.36 [8.72]& 1.66 [0.18]& 0.64 [0.17]& 0.70 [0.25]\\
fML1000\_nu1e10 & $10^{3}$ & $10^{10}$ & 6.08 [0.52]& 0.28 [0.14]& 498  & 2.57 [5.14]& 0.22 [0.02]& 0.13 [0.03]& 0.16 [0.06]\\    
\hline  
\end{tabular}
\end{table*}

In calculating the retained energy fraction, we note that it is
important to include both the mechanical energy input by the wind {\em
  and} the integrated thermal energy of the ambient gas that has been
swept up (which becomes significant at late times in this model). The
energy input by the wind $E_{\rm w} = \dot{E}t$, where
$\dot{E}=\frac{1}{2}\Mdot_{\rm w}v_{\rm w}^{2}$ is the mechanical
luminosity of the wind. The thermal energy of the swept-up gas is
$E_{\rm sw} = 1.5\,V_{\rm bub}P_{\rm ic}$. The total input energy is
$E_{\rm i} = E_{\rm w} + E_{\rm sw}$. Fig.~\ref{fig:wbb_fML0}h) shows
that the bubble retains about 75 per cent of the input energy at
$t\approx0.5$\,Myr, but just 53 per cent at $t=5$\,Myr (at this time,
$E_{\rm w} = 2\times10^{49}\,{\rm erg}$,
$E_{\rm sw} = 1.3\times10^{49}\,{\rm erg}$, and the total energy
measured in the bubble is $E=1.75\times10^{49}\,{\rm erg}$).

In the case of an ideal adiabatic bubble expanding into a
pressure-less environment, $E_{\rm i}=E_{\rm w}$ since
$E_{\rm sw}=0.0$. In such cases we expect the swept-up shell to have a
kinetic energy of $\frac{15}{77}E_{\rm w}$
\citep[e.g.][]{Dyson:1980}. Behind a strong shock, the kinetic energy
and thermal energy per unit mass are identical, so up to 19 per cent
of the input energy can be radiated by the swept-up gas. This fraction
is in rough agreement with the roughly 25 per cent energy loss seen at
$t\approx0.5$\,Myr, with cooling in the hot bubble and at the contact
discontinuity adding the remainder. The decrease seen in the retained
energy fraction between 0.5 and 5\,Myr indicates that cooling of the
shocked stellar wind becomes more significant as the bubble
ages. Having said this, the retained energy fraction reaches a minimum
near $t=5\,$Myr and then starts to rise slightly. This is due to the
increasing significance of the thermal energy of the swept-up gas, and
the fact that this gas suffers little radiative loss at late times
since it is heated very little and the post-shock temperature remains
close to the equilibrium temperature for gas at such densities. In any
case, the retained energy fraction is always above 50 per cent, and
the bubble behaviour indicates that radiative energy losses from the
hot gas in the bubble interior have little consequence in this model.

In summary, we find that without mass-loading the bubble expands as
expected given that the ambient pressure becomes significant at late
times. The bubble does significant $PdV$ work on the surrounding gas,
boosting the radial momentum input by the wind by a factor of 40 by
$t=5$\,Myr.

\subsubsection{A large reservoir of clump mass}
We now examine the effect of mass-loading on a WBB. We begin by
assuming that there is an effectively infinite reservoir of mass in
the clumps, which never runs out. We achieve this in the simulations
by setting $\nu$ to a very high value ($\nu=10^{10}$). We explore
mass-loading factors of $10^{2-3}$. This is motivated by estimated
factors of $\approx 170$ in the dusty wind-blown bubble N49
\citep{Everett:2010}, $40-50$ in the Wolf-Rayet wind-blown-bubble
RCW~58 \citep{Smith:1984} and $\sim100$ in the halo of the core-halo
planetary nebula NGC\,6543 \citep*{Meaburn:1991,Arthur:1994}, as well
as factors of up to several hundred occuring in numerical simulations
\citep{Rogers:2013}.

\begin{figure*}
\includegraphics[width=16.0cm]{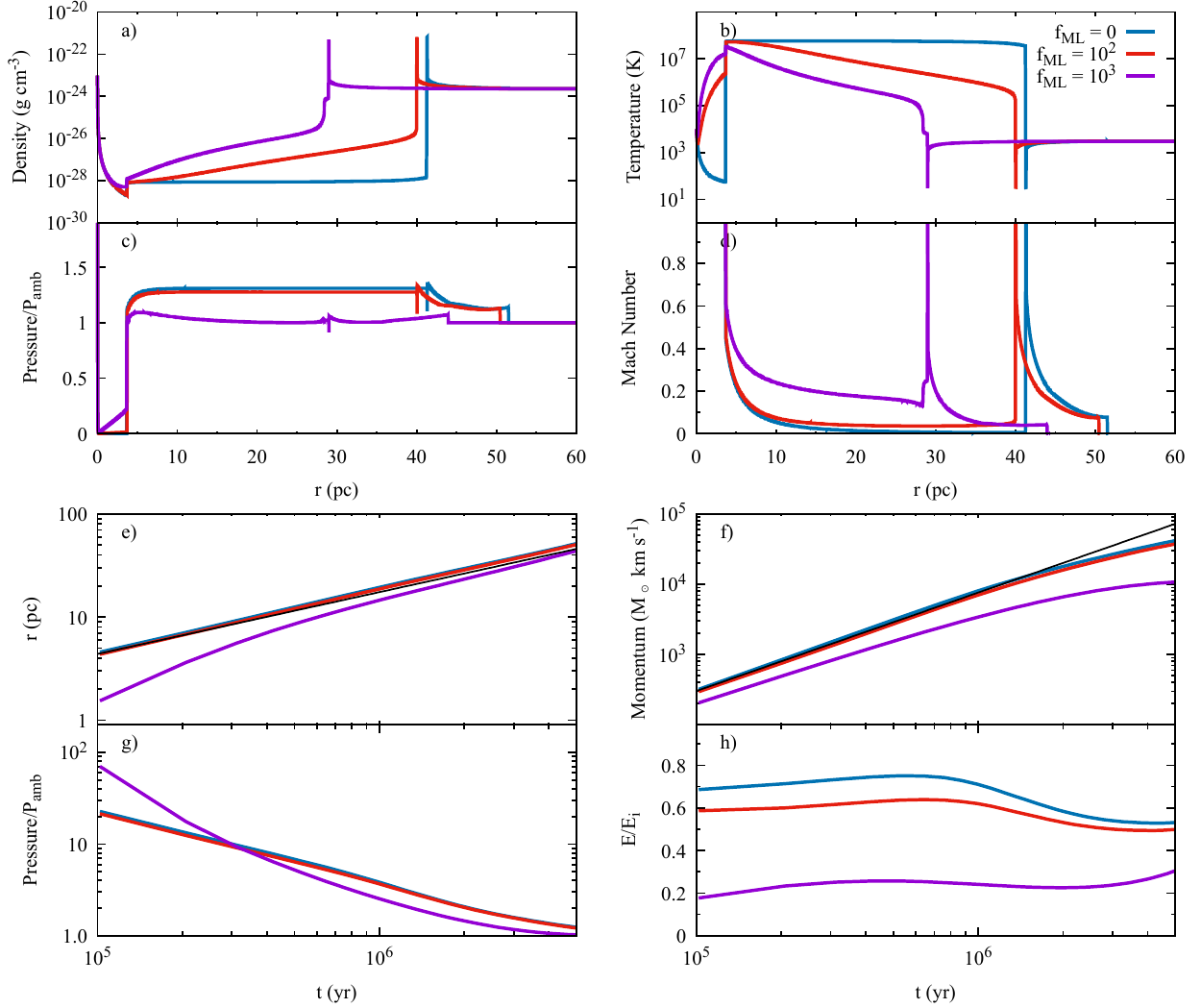}
\caption{The evolution of a bubble with
  $n_{\rm H,ic}=1\,{\rm cm^{-3}}$
  ($\rho_{\rm ic}=2.267\times10^{-24}\,{\rm g\,cm^{-3}}$) and with
  varying amounts of mass-loading. Strong mass-loading (high values of
  $f_{\rm ML}$) results in dramatic changes in the bubble
  properties. The profiles in panels a)-d) are at $t=5$\,Myr.}
\label{fig:wbb_varfML}
\end{figure*}

In Fig.~\ref{fig:wbb_varfML} we show profiles of the WBB at
$t=5$\,Myr, as a function of the strength of mass-loading.  Some
dramatic differences are visible in the profiles when mass-loading
from embedded clumps occurs. When $f_{\rm ML}=10^{2}$ we see that the
mass-loading increases the density and decreases the temperature of
the shocked stellar wind gas. The latter is mainly due to sharing the
thermal energy of the gas between more particles. We see also that
mass-loading of the unshocked wind increases its temperature prior to
passing through the reverse shock. This is due to the frictional
aspect of mass-loading. The density of this part of the flow also
increases but the change is minimal in this model. We also find that
the amount of mass added to a particular part of the shocked wind
increases with distance from the reverse shock. This is because the
oldest stellar wind material (defined as the time since being emitted
from the star) is closest to the contact discontinuity. At the reverse
shock the fraction of injected mass is about 4 per cent, while it
increases to 99.1 per cent of the gas mass at the contact
discontinuity. Mass-loading also occurs in the thick swept-up shell,
but the injected mass fraction in this region remains below 2 per
cent.

Fig.~\ref{fig:wbb_varfML}c) shows that mass-loading has reduced the
pressure of the bubble at this time, although as can be seen in
Fig.~\ref{fig:wbb_varfML}g) this is not necessarily true at earlier
times for bubbles with strong mass-loading. Mass-loading also causes
an increase in the Mach number of the shocked stellar wind, as
expected \citep*[see][]{Hartquist:1986,Arthur:1993,Arthur:1996}.

From Table~\ref{tab:ML_results} and Fig.~\ref{fig:wbb_varfML}, we see
that with $f_{\rm ML}=10^{2}$ there are only minor differences in the
radius, total energy and radial momentum of the bubble, indicating
that the density and temperature changes within the bubble caused by
the mass-loading have not resulted in significant additional radiative
cooling. However, the mass of hot gas has increased from $0.54\,\Msol$
to $17.6\,\Msol$. Thus, while in this case mass-loading has not
dramatically changed the dynamics of the bubble (e.g., forward shock
radius, total radial momentum, etc.), it will have significantly
affected its X-ray emission.

\begin{figure*}
\includegraphics[width=16.0cm]{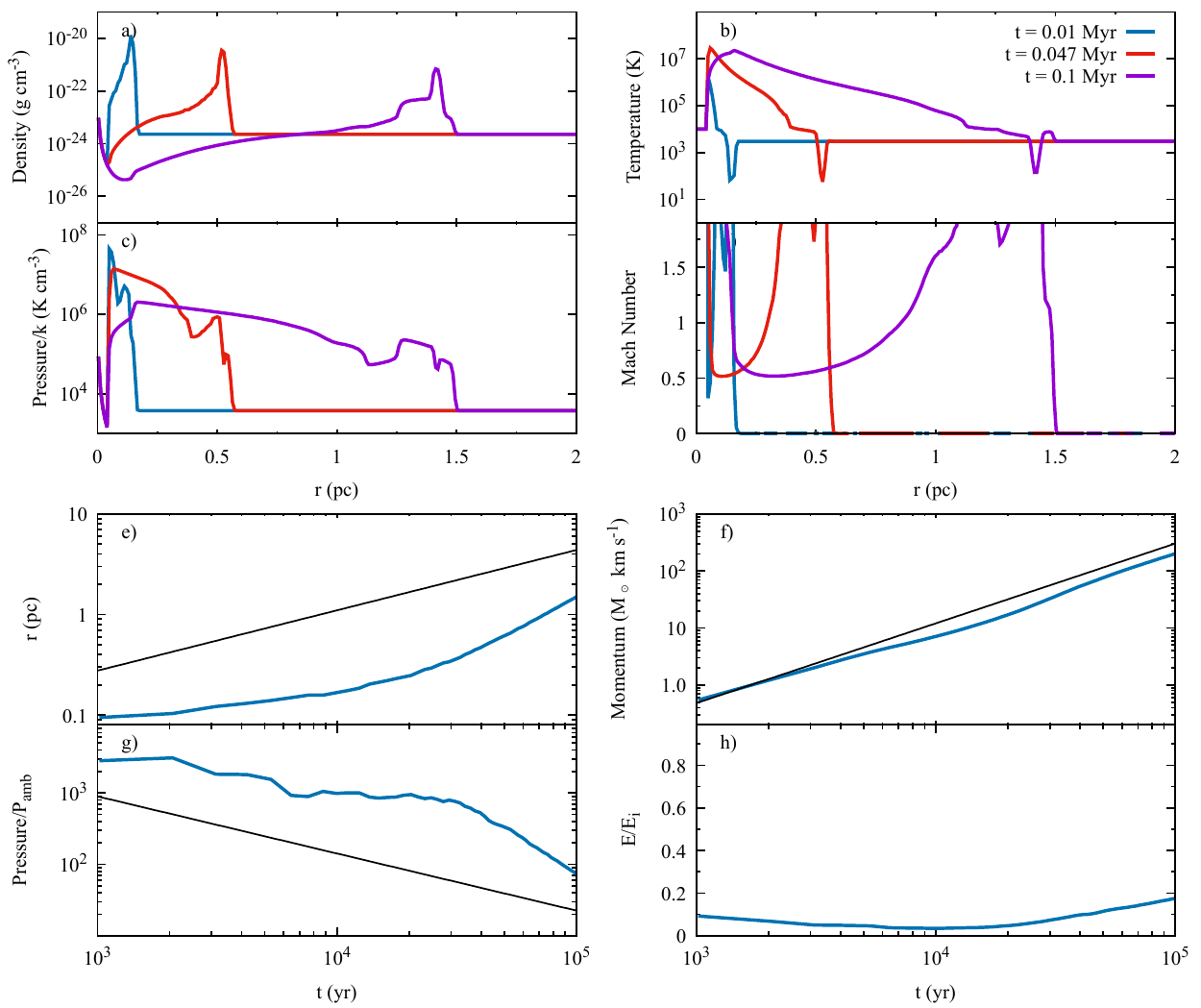}
\caption{As Fig.~\ref{fig:wbb_fML0} but for a bubble with strong
  mass-loading ($f_{\rm ML}=10^{3}$), and with a focus on its early ($t
  \leq 0.1$\,Myr) development.}
\label{fig:wbb_fML1e3_early}
\end{figure*}

\begin{figure*}
\includegraphics[width=16.0cm]{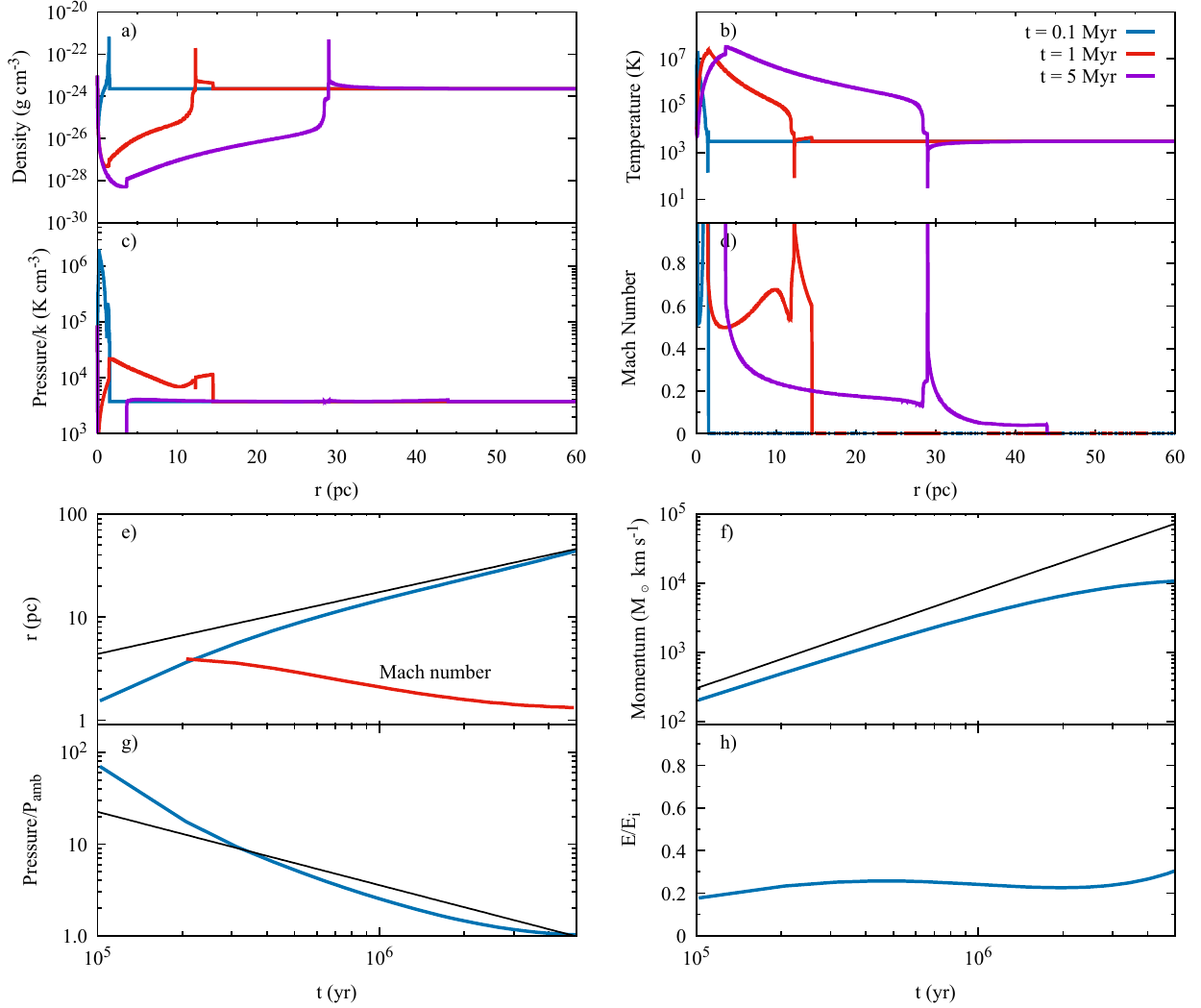}
\caption{As Fig.~\ref{fig:wbb_fML1e3_early} but focusing on the later ($t
  \geq 0.1$\,Myr)
  development of the bubble.}
\label{fig:wbb_fML1e3_late}
\end{figure*}

There are much more significant differences when $f_{\rm ML}=10^{3}$,
the most notable being that the bubble is much smaller at early times,
and the radial momentum and retained energy fraction are also much
smaller than the standard bubble.  To better understand this
difference in behaviour, we show in Figs.~\ref{fig:wbb_fML1e3_early}
and~\ref{fig:wbb_fML1e3_late} the early and late evolution of the
$f_{\rm ML}=10^{3}$ bubble.

We see from Fig.~\ref{fig:wbb_fML1e3_early}b) that at $t=0.01$\,Myr
the hottest gas has a temperature of about $10^{6}$\,K, which is far
below the temperature of the bubble with no mass-loading. There is no
evidence of a reverse shock at this time, and the mass-loading has an
immediate and strong effect on the flow in the first grid cell outside
of the wind injection region, to the extent that 97 per cent of the
mass in this cell is injected from the clumps, and the velocity is
slowed to 3 per cent of the wind speed. However, the wind is able to
slowly push away and by $t=0.047$\,Myr, the wind maintains 71 per cent
of its initial speed and makes up 81 per cent of the mass in the first
grid cell outside of the wind injection region. However, the density
in this first cell still exceeds the density in the final grid cell
inside the wind injection region at this time and there is still no
sign of a reverse shock. Finally, by $t=0.1$\,Myr, a (very weak)
reverse shock is established at $r=0.15$\,pc (this is best seen in the
pressure jump in Fig.~\ref{fig:wbb_fML1e3_early}c).

\citet{Pittard:2021} showed that simulations of wind-blown bubbles
must have sufficient resolution such that the reverse shock initially
moves away from the edge of the wind injection region, otherwise the
amount of $PdV$ work done by the bubble will be underestimated. We
have ensured that we have enough resolution in our calculations
without mass-loading to meet this requirement, but it is clear that it
is not fully met in our $f_{\rm ML}=10^{3}$ simulations. However, we
can confirm that in higher resolution calculations, a weak reverse
shock is established between $2000-3000$\,yr. Although there are some
quantitative differences in the profiles and bubble properties at
$t=10^{5}$\,yr, these differences reduce with time (e.g. the
difference in the radial momentum at $t>0.5$\,Myr is less than 5 per
cent). Likewise, investigation of the $f_{\rm ML}=10^{2}$ simulations
reveals that the reverse shock is established in the standard
resolution calculations by $t=3100$\,yr. Thus we are confident that
all of our models are capturing the physics correctly and are accurate
enough for our purposes.

So why does the $f_{\rm ML}=10^{3}$ bubble evolve so differently to
bubbles with $f_{\rm ML}=0$ and $10^{2}$? It is clear that the more
highly mass-loaded bubble radiates significantly more energy, with the
retained energy fraction falling to about 2 per cent at
$t=10^{4}$\,yr. Thus it appears that the added mass in the bubble
causes the bubble to cross a threshold where cooling is finally able
to become significant in the bubble interior\footnote{The effect of
  strong cooling when bubbles become too mass-loaded has been studied
  in the case of super-star cluster winds by \citet{Wunsch:2011} and
  \citet{Silich:2013}.}. This changes the interior pressure (though
not in a simple way due to the frictional effect of the mass-loading),
and slows the bubble expansion. Velocities in the bubble are also
reduced due to the necessity for the flow and injected mass to
conserve momentum.

As the $f_{\rm ML}=10^{3}$ bubble expands, the density of the hot
interior decreases, and the cooling becomes somewhat less effective,
causing the retained energy fraction to increase to a value of about
0.25 for most of the bubble lifetime. Analysis of the cooling in this
model indicates that roughly 25 per cent is by gas with
$T > 10^{5}$\,K. Concerning the origin of the gas causing the cooling,
22 per cent is from swept-up ambient material, and 78 per cent is from
clump material (as indicated by the value of the passive scalar). The
clump material is mostly mixed in with the shocked stellar wind and,
although it exists at a wide range of temperatures, predominantly
cools at $T < 10^{5}$\,K. Nevertheless, the bubble still performs
significant $PdV$ work during its life, boosting the momentum input by
the wind by a factor of 10.

In summary, mass-loading reduces the retained energy in the bubble
(see Fig.~\ref{fig:wbb_varfML}h and Table~\ref{tab:ML_results}), but
hot gas is still present in our models which allows the bubble to
still do significant $PdV$ work.

\subsubsection{A finite amount of clump mass}
\begin{figure*}
\includegraphics[width=16.0cm]{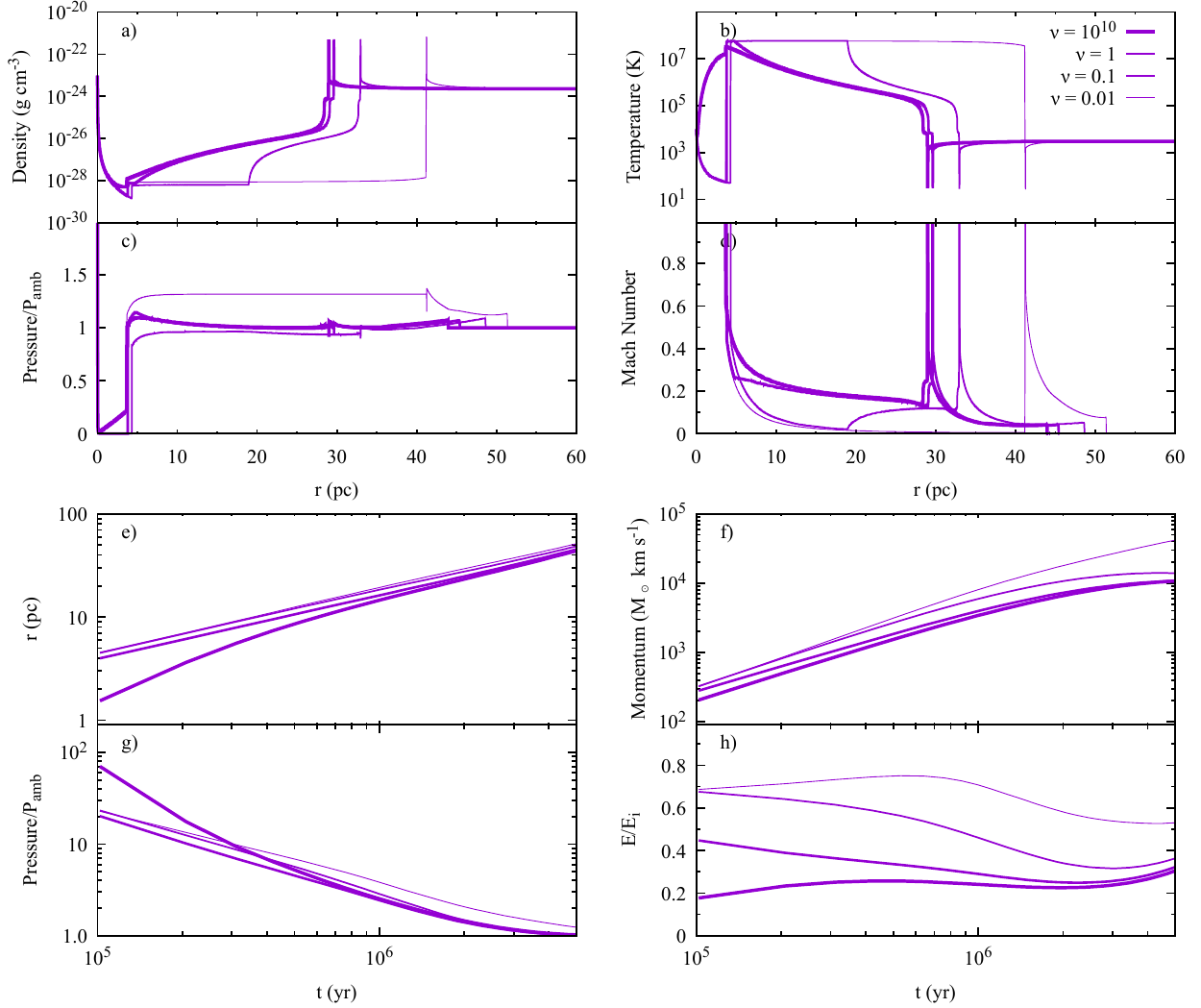}
\caption{Results from a model with $n_{\rm H,ic}=1 \,{\rm cm^{-3}}$
  ($\rho_{\rm ic}=2.267\times10^{-24}\,{\rm g\,cm^{-3}}$),
  $f_{\rm ML}=10^{3}$ and $\nu = 10^{10}, 1.0, 0.1$ and 0.01. As $\nu$
  decreases the amount of mass available in the clumps that can be
  injected into the bubble reduces. The profiles in panels a)-d) are
  at $t=5$\,Myr.}
\label{fig:wbb_finitereservoir}
\end{figure*}

We now investigate the behaviour of a WBB subject to rapid
mass-loading ($f_{\rm ML}=10^{3}$) but where there is a finite amount
of available mass in the
clumps. Fig.~\ref{fig:wbb_finitereservoir}a)-d) shows density,
temperature, pressure and Mach number profiles in this case. We see
that when there is equal mass in the clumps and intercloud gas
(i.e. $\nu=1$), the WBB still shows the effects of significant
mass-loading. This is because the average smeared out density of the
clumps, which is equal to the ambient intercloud density, is
significantly greater than the density of the shocked wind in the
bubble interior, and therefore the injected mass can still
dramatically reduce the temperature of the bubble interior.

\begin{figure}
\includegraphics[width=8.0cm]{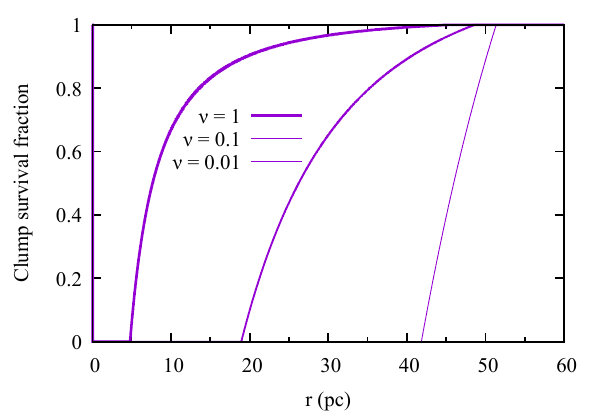}
\caption{Profiles of the clump survival fraction at $t=5$\,Myr from
  simulations with $n_{\rm H,ic}=1 \,{\rm cm^{-3}}$
  ($\rho_{\rm ic}=2.267\times10^{-24}\,{\rm g\,cm^{-3}}$),
  $f_{\rm ML}=10^{3}$ and $\nu = 1.0, 0.1$ and 0.01. As $\nu$
  decreases the region over which all clumps have been destroyed
  expands to larger radii.}
\label{fig:clump_survival}
\end{figure}

However, when only 1 per cent of the background mass is in clumps
(i.e. $\nu=0.01$), Fig.~\ref{fig:wbb_finitereservoir} shows that there
is insufficient mass in the clumps to significantly affect the bubble
interior, which now resembles that of a bubble without any
mass-loading. Fig.~\ref{fig:clump_survival} shows the clump survival
fraction for these simulations, which we define as the fraction of the
initial clump mass that still remains (i.e. the current value of
$\rho_{\rm cl,avg}$ divided by its initial value). When $\nu=0.01$, we
see that clumps are only present in the region of swept-up gas, and
that interior to the contact discontinuity no clumps survive. Thus the
ongoing mass-loading actually occurs only in the swept-up
shell. Clumps that are overrun by the bubble do not survive their
passage through the thick shell, and none reach the low density, hot
interior gas.

Most interestingly, we see that the $\nu=0.1$ simulation represents an
intermediate stage where the clumps have enough total mass to
significantly affect the density, temperature, pressure and Mach
number of the bubble interior, but not enough mass to ensure that
mass-loading continues in all parts of the bubble over its entire
life. In this case, at $t=5$\,Myr, no clumps survive at $r <
19$\,pc. This arises from the fact that the clouds that are closest to
the central star interact with the bubble at earlier times than more
distant clouds, and the mass injected from the clouds is swept
downstream towards the edge of the bubble\footnote{In reality the
  clouds will likely pick up some momentum while they are being
  destroyed.}. This has an interesting effect on the $\nu=0.1$
profiles shown in Fig.~\ref{fig:wbb_finitereservoir}. We see a
significant density enhancement only for $r > 19$\,pc, where the
temperature rapidly drops and the Mach number climbs as mass-loading
continues from the parts of the mass reservoir that are as yet
un-depleted. However, between the reverse shock at $r=4.3$\,pc and
$r=19$\,pc, the temperature of the gas is $\approx5\times10^{7}$\,K
(essentially that of a bubble without any mass-loading).

In Fig.~\ref{fig:wbb_finitereservoir}e)-h) we show the radius, radial
momentum, pressure and retained energy of the bubble as a function of
time. We see that the radius of the forward shock in the model with
$\nu=0.1$ at first diverges from the $\nu=0.01$ case, but after
$2-3$\,Myr it begins to converge again as the relative lack of clump
mass begins to be felt. Fig.~\ref{fig:wbb_finitereservoir}f) reveals
that when $\nu=0.1$ the radial momentum plateaus at late
times. Clearly the mass-loading at early times for the $\nu=0.1$ case
is somewhat constrained by the available reservoir of cloud mass (note
that the momentum is initially much closer to the $\nu=0.01$ case than
the $\nu=1$ case), but by $t=1$\,Myr a significant momentum difference
has arisen between the $\nu=0.01$ and $\nu=0.1$ simulations. This
difference increases with time until the end of the simulations, so
that a significant reduction in the final momentum still occurs when
$\nu=0.1$.

Fig.~\ref{fig:wbb_finitereservoir}g) shows that the bubble pressure
responds in a non-linear way to changes in $\nu$. Within the range
$\nu=0.01-1$, increasing $\nu$ leads to a reduced bubble pressure at
all times. However, when there is an infinite amount of clump mass,
the bubble pressure may be greater or smaller than a bubble without
mass loading, depending on the bubble age. On the other hand,
Fig.~\ref{fig:wbb_finitereservoir}h) shows that the retained energy
fraction varies in a more straightforward way - the less mass-loading,
the higher the retained energy.

\begin{figure*}
\includegraphics[width=16.0cm]{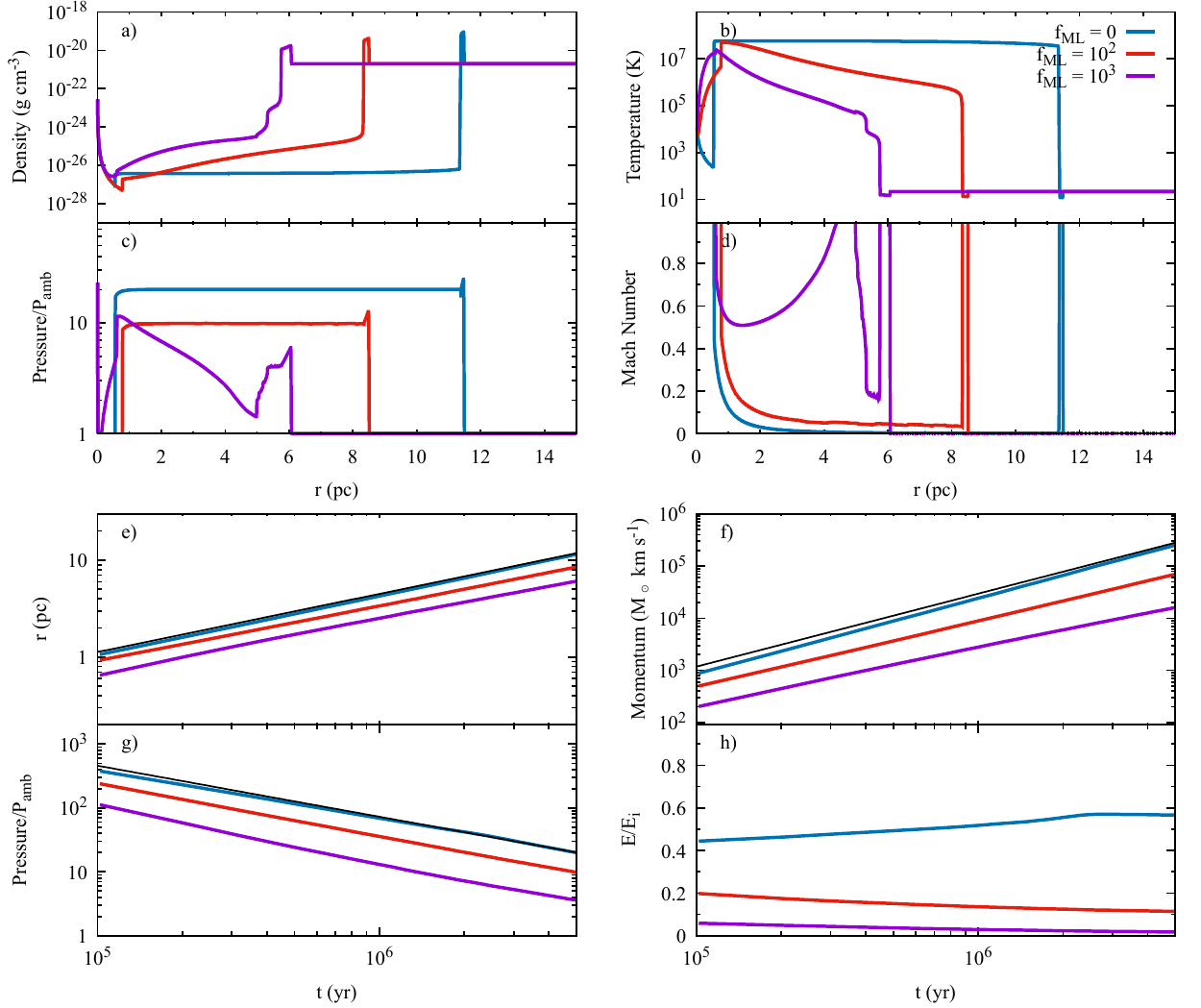}
\caption{As Fig.~\ref{fig:wbb_varfML} but for an intercloud density
  $n_{\rm H,ic}=884\,{\rm cm^{-3}}$
  ($\rho_{\rm ic}=2\times10^{-21}\,{\rm g\,cm^{-3}}$). The profiles in
panels a)-d) are at $t=5$\,Myr.}
\label{fig:higherramb}
\end{figure*}

\subsection{``High'' intercloud density}
We have also investigated the evolution of WBBs in a denser
environment. Specifically we set $n_{\rm H,ic}=884\,{\rm cm^{-3}}$
($\rho_{\rm ic} = 2.0\times10^{-21}\,{\rm g\,cm^{-3}})$, which gives
$T_{\rm ic} = 21.2$\,K. The pressure of the intercloud gas,
$P_{\rm ic} = 1.48\times10^{-12}\,{\rm dyn \,cm^{-2}}$
($P_{\rm ic}/k = 1.075\times10^{4}\,{\rm K \,cm^{-3}}$). With these
parameters $r_{\rm inj,max} = 2.68$\,pc. The width of each grid cell
is set to $dr=5.36\times10^{-3}$\,pc. Fig.~\ref{fig:higherramb} and
Table~\ref{tab:ML_results_ramb2e-21} show the results for this
scenario.

The main difference to the lower density simulations is that the
bubble is much smaller at any given time. This results in much higher
pressures inside the bubble, and although the ambient pressure is
nearly 3 times greater, higher relative pressures in the bubble mean
that the ambient pressure remains negligible throughout the
simulation: hence the simulation without mass-loading agrees well with
simple analytical theory even in the latter stages of the bubble's
evolution.

Examining first the $f_{\rm ML}=0$ bubble without any mass-loading, we
note that the isothermal Mach number of the forward shock remains high
throughout the simulation (at $t=5\,$Myr it has a value of 5.1), which
causes the swept-up shell to remain thin - the compression at the
shell is about a factor of 45. This compression is greater than the
factor of 26 expected from the isothermal Mach number, and the
difference arises because the gas doesn't remain isothermal.  Instead,
the gas temperature decreases from the intercloud ambient temperature
of 21.2\,K to 11.7\,K.

The smaller bubble size leads to significantly higher densities in the
shocked stellar wind. The smaller bubble size means that the thermal
energy of the swept-up ambient medium
($E_{\rm sw}=6.9\times10^{47}\,{\rm erg}$) does not add significantly
to the total energy of the bubble. Without any mass-loading, we find
that the bubble has radiated about 55 per cent of its energy at
$t=0.1$\,Myr (significantly more at this stage than the bubble
expanding into the lower density environment). The retained energy
fraction increases with time as the bubble expands and the density of
the shocked wind drops. At $t=5$\,Myr, 42 per cent of the input energy
has been radiated away (19 per cent by the swept up gas, and 23 per
cent by the shocked wind). Nevertheless, the bubble interior is hot
and the bubble is actually able to do more $PdV$ work than the
equivalent bubble expanding into a lower density medium, with the
momentum boost reaching a factor of 250.

We also see that mass-loading seems to have more of an effect on the
bubble radius and momentum for a given value of $f_{\rm ML}$. There is
now a significant difference between the $f_{\rm ML}=0$ and
$f_{\rm ML}=10^{2}$ models, whereas the differences were minimal when
$n_{\rm H,ic}=1\,{\rm cm^{-3}}$ (see Fig.~\ref{fig:wbb_varfML}). This
seems to be because of the stronger cooling in the shocked stellar
wind, which causes the retained energy fraction to drop to 12 and 2
per cent for $f_{\rm ML}=10^{2}$ and $10^{3}$, respectively. Thus the
threshold mass-loading factor where significant additional radiative
cooling occurs is somewhat lower when the intercloud ambient density
is higher and the bubble relatively smaller and denser. It is also
interesting to see that the retained energy fraction increases or is
constant with time for the bubble without mass-loading, but decreases
with time for the mass-loaded bubbles.

The final radial momentum in the $f_{\rm ML}=10^{3}$ simulation
now decreases to 6 per cent of the value obtained from the
equivalent bubble without any mass-loading, though this
still represents a factor of 16 boost over the input wind momentum.

\section{Discussion}
\label{sec:discussion}

\subsection{Validity of the assumptions}
A key assumption in this work is that the clouds mix rapidly with the
background flow, with negligible radiative losses during this
time. The interaction of the hot intercloud gas with the cooler cloud
material creates a turbulent mixing layer with a characteristic
temperature $T_{\rm m} = \sqrt{T_{\rm h}T_{\rm c}}$, where $T_{\rm h}$
and $T_{\rm c}$ are the temperatures of the hot and cold gas,
respectively \citep{Begelman:1990}. The characteristic number density
of the mixing layer
$n_{\rm m} = n_{\rm h}\sqrt{T_{\rm h}/T_{\rm c}} = n_{\rm
  c}\sqrt{T_{\rm c}/T_{\rm h}}$, where $n_{\rm h}$ and $n_{\rm c}$ are
the number densities of the hot and cold gas, respectively. An
excellent review of the nature of such interfaces is given by
\citet{Hartquist:1988}.

Recent numerical studies by \citet{Gronke:2018} showed that clouds in a hot
wind are destroyed if
\begin{equation}
  \label{eq:tcc_vs_tcoolmix}
  t_{\rm cc} < t_{\rm cool,mix},
\end{equation}
where $t_{\rm cc}$ is the cloud crushing time, and $t_{\rm cool,mix}$
is the cooling timescale of the mixed
gas. Eq.~\ref{eq:tcc_vs_tcoolmix} sets an upper limit to the size of
the clouds, since $t_{\rm cc} \propto r_{\rm c}$, where $r_{\rm c}$ is
the cloud size or radius\footnote{If the opposite is true
  ($t_{\rm cool,mix} < t_{\rm cc}$), hot gas can condense onto the
  cloud and the cloud can gain mass \citep{Gronke:2018,Gronke:2020}.}.

The cooling time of the mixed gas,
\begin{equation}
  t_{\rm cool,mix} = \frac{E}{\dot{E}} = \frac{3\,n_{\rm m}kT_{\rm
      m}}{2 n_{\rm m}^{2}\Lambda(T_{\rm m})}. 
\end{equation}
Given that $t_{\rm cc} = \chi^{1/2} r_{\rm c}/v_{\rm rel}$, where $v_{\rm rel}$
if the relative velocity between the hot and cold gas, for the cloud
to be shredded ($t_{\rm cc} < t_{\rm cool,mix}$), we require
that
\begin{equation}
\label{eq:rcloud_max}  
r_{\rm c} < 1.5  \frac{v_{\rm rel}kT_{\rm m}}{\sqrt{\chi} n_{\rm m}\Lambda(T_{\rm m})}.
\end{equation}  
The value of $r_{\rm c}$ from Eq.~\ref{eq:rcloud_max} is highly
dependent on the temperature of the mixed gas since $\Lambda(T)$ rises
so steeply around $10^{4}$\,K.

Let us assume that the clouds have a temperature of about $10$\,K and
the hot gas in the bubble is at a temperature of about
$10^{7}$\,K. This gives a temperature for the mixing layer of about
$10^{4}$\,K. With our cooling curve,
$\Lambda(T)=10^{-24}\,{\rm erg\,cm^{3}\,s^{-1}}$ when $T=8700$\,K. We
will therefore use this value in the following calculations, but note
that $\Lambda(T)$ is two orders of magnitude lower at $T=1600$\,K and
two orders of magnitude higher at $T=15,500$\,K. The maximum cloud
radius that satisfies Eq.~\ref{eq:tcc_vs_tcoolmix} will therefore be
much smaller (larger) than our estimate if $T_{\rm m}$ is only
slightly higher (lower) than $T=8700$\,K.

We continue by noting that the WBB has a typical overpressure relative
to the ambient medium of $10-100\times$ (see, e.g.,
Figs.~\ref{fig:wbb_fML0}c,~\ref{fig:wbb_varfML}c
and~\ref{fig:higherramb}c). Taking
$P_{\rm amb}/k \sim 10^{4}\,{\rm K\,cm^{-3}}$, this means that the
embedded clouds have a pressure
$P_{\rm c}/k \sim 10^{5-6}\,{\rm K\,cm^{-3}}$. If $T_{\rm c}=10$\,K,
then $n_{\rm c} \sim 10^{4}-10^{5}\,{\rm cm^{-3}}$. In the following
calculations we assume that $n_{\rm c} = 10^{4}\,{\rm cm^{-3}}$.  With
$T_{\rm h}\approx 10^{7}$\,K, we obtain
$\chi=T_{\rm h}/T_{\rm c} = n_{\rm c}/n_{\rm h} = 10^{6}$ and
$n_{\rm h} = 10^{-2}\,{\rm cm^{-3}}$. This gives
$T_{\rm m} \approx 10^{4}$\,K (we use 8700\,K in our calculations) and
$n_{\rm m} = 10\,{\rm cm^{-3}}$.

If $v_{\rm rel} = v_{\rm w}/4 = 500\,{\rm km\,s^{-1}}$ (this is
typical of the flow speed just after the reverse shock, but the gas
slows as it moves towards the contact discontinuity) we obtain
$t_{\rm cool,mix}=5700$\,yr and $r_{\rm c} < 3\times10^{-3}$\,pc.
Clouds of this size and smaller, with similar densities, have been
observed in many \hii regions
\citep[e.g.][]{deMarco:2006,Gahm:2013,Grenman:2014,Haikala:2017}.  The
maximum mass of the cloud is $m_{\rm c} = 0.03\,M_{\rm J}$ (Jupiter
masses). Smaller values of $v_{\rm rel}$ mean longer values of
$t_{\rm cc}$ which requires smaller clouds to satisfy
Eq.~\ref{eq:tcc_vs_tcoolmix}. The cooling time of the gas in the
mixing layer is similar to the cooling times seen in Figs.~2 and~3 of
\citet{Lancaster:2021b}.

In the $f_{\rm ML} = 10^{3}$ simulations with
$n_{\rm ic}=1\,{\rm cm^{-3}}$, $500\,\Msol$ of cloud material was
injected into the bubble by $t=5$\,Myr. This corresponds to the
destruction of more than $1.8\times10^{7}$ clumps, and since the WBB
has a radius of 44\,pc, the clumps have a volume filling factor of
$\sim 10^{-5}$ and thus take up a negligible amount of space within
the bubble. The ratio of the total surface area of the clumps to the
surface area of the WBB is $\gtrsim 0.1$.

Directly modelling such a range of length scales is impossible with
current computational resources. Ideally one would like to have a
resolution, $\Delta x$, such that there are of order 100 grid cells
per cloud radius. This requires that
$\Delta x \lesssim 3\times10^{-5}$\,pc. Capturing the global WBB at
the same time as resolving the interaction around individual clouds
requires $\gtrsim 10^{6}$ cells per grid axis, or $\gtrsim 10^{18}$ 3D
grid cells. Relaxing the resolution requirements to 10 cells per cloud
radius, as suggested by \citet{Banda-Barragan:2020}, and modelling
just one octant requires $\gtrsim 10^{14}$ cells. Due to the turbulent
nature of the flow, it is unlikely that adaptive mesh refinement will
be of much use. Therefore, our approach of treating the clumps as a
contiuous distribution is the only feasible method for simulating WBBs
that are mass-loaded by small clouds at the current time.

Finally, we note that due to a lack of significant bulk motions in IFU
observations of a gas pillar in the \hii region NGC~6357,
\citet{Westmoquette:2010} conjectured that the evaporated and/or
ablated gas from the pillar is rapidly heated before it is mixed
and/or entrained into the surrounding flow. This provides some
observational support for our assumption that the mass injected from
clumps into our bubbles does not undergo significant radiative cooling
during this process.

We can also wonder what effect the ionizing photons from the central
star may have on the clumps\footnote{In reality, the stellar wind will
  always interact with a \hii region where clumps are subject to the
  ``rocket effect'' which homogenizes the region that the wind is
  interacting with \citep*[][]{Elmegreen:1976,McKee:1984}.}. A star
with similar wind properties has a hydrogen ionizing photon flux
$Q_{\rm H} \approx 10^{48}\,{\rm s^{-1}}$
\citep*{Sternberg:2003}. Using the equations in \citet{Bertoldi:1989},
we find that clouds with $r_{\rm c}=3\times10^{-3}$\,pc and
$n_{\rm c} = 10^{4}\,{\rm cm^{-3}}$ will be instantly ionized (or
``zapped'') if closer than $0.13$\,pc. At a distance of 5\,pc from the
star, we find that the clouds lie in region II of Fig.~1 in
\citet{Bertoldi:1989} and so will be compressed by an ionization shock
front which is thin compared to the size of the cloud. The ionized gas
flows away from the cloud, causing the cloud to lose mass at a rate
$\Mdot_{\rm c,ph} = mFA$, where $m$ is the mass per particle of the
neutral material, and $F$ and $A$ are the rate per unit time per unit
area at which hydrogen ionizing photons reach the ionization front and
its area, respectively \citep{Mellema:1998}. To first order, the
lifetime of the clump before it is completely photoevaporated is
$t_{\rm life,ph} = M_{\rm c}/\Mdot_{\rm c,ph}$ (this is a lower limit
since in reality the mass-loss rate decreases with time). We estimate
that $F = F_{0}/1.1$, where $F_{0}=Q_{\rm H}/4\pi d^{2}$, giving
$\Mdot_{\rm c,ph}=1.3\times10^{17}\,{\rm g\,s^{-1}}$. The cloud will
then have a lifetime of approximately $1.3\times10^{4}$\,yr. Much
smaller clouds will lie in region V, whereby the
ionization-front-driven shock sweeps rapidly over the cloud. Clouds
which are further from the star will have lower rates of
photoevaporation (smaller $\Mdot_{\rm c,ph}$) and longer
lifetimes. Since the cloud crushing time, $t_{\rm cc}=5700$\,yr,
$t_{\rm life,ph}$ and the cloud mixing time, $t_{\rm mix}$, are of
similar magnitude\footnote{In adiabatic hydrodynamic simulations,
  $t_{\rm mix}\approx 5-15\,t_{\rm cc}$
  \citep{Pittard:2016a,Pittard:2016b}.}. Hence, in reality both
hydrodynamic ablation and photoevaporation likely play a role in
destroying the clouds.

To summarize, the requirement that the cloud mixing time be less than
the cooling time of the mixed gas requires that our embedded clouds
have radii $r_{\rm c} \lesssim 3\times10^{-3}$\,pc. Significantly
smaller clouds may be immediately zapped by the ionizing radiation
from the central star. Clouds at this size limit will instead have a
lifetime against photoevaporation of order
$1.3\times10^{4}$\,yr. Clouds may also lose mass through thermal
conduction \citep[see][for a discussion of these different
mechanisms]{Pittard:2007}. Our numerical values, are, however, very
sensitive to the temperature of the gas in the mixing layer. If this
is slightly lower than we have assumed, much larger clouds can be
destroyed, and fewer clouds are needed to provide the required mass
injection into the bubble. This could easily arise in situations where
the temperature of the hot gas is a little lower than we have assumed
(e.g. if $v_{\rm w} < 2000\kmps$, or if the cloud is interacting with
part of the flow that has already experienced some
mass-loading). However, if $T_{\rm m}$ is only slightly higher than we
have assumed, the hot-phase gas will instead try to condense onto the
cold clouds (although in this case the clouds likely still lose mass
due to photoevaporation). Irrespective of how the clouds lose mass,
the mass injected into the surrounding flow can have a significant
effect on the global properties of the flow, as this work shows.

\subsection{Comparison to previous mass-loading simulations}
Similarity solutions of mass-loaded WBBs were obtained by
\citet*{Pittard:2001a} and \citet*{Pittard:2001b}. They found that
with extremely high mass-loading the wind could be slowed to such an
extent that it connects directly to the contact discontinuity, without
the presence of a reverse shock. Although we find that vigorous
mass-loading slows the creation of a reverse shock (see
Fig.~\ref{fig:wbb_fML1e3_early}), all of our models have a reverse
shock by $t=0.1$\,Myr. A reverse shock still forms when
$f_{\rm ML} = 10^{4}$, but when $f_{\rm ML} = 10^{5}$ the reverse
shock completely disappears and is not present at $t=5$\,Myr. We also
find that for $f_{\rm ML}=10^{4}$ and $10^{5}$, the final radial
momentum slightly {\em exceeds} that obtained when $f_{\rm ML}=10^{3}$
(at $t > 3$\,Myr; before this time the radial momentum is
lower)\footnote{The $f_{\rm ML} = 10^{5}$ bubble at $t=5$\,Myr has
  some similarities to the $f_{\rm ML} = 10^{3}$ bubble at very early
  times (see Fig.~\ref{fig:wbb_fML1e3_early}). Only a small amount of
  gas near the center of the bubble where the stellar wind starts
  being mass-loaded is hot. This gas rapidly heats due to the
  frictional heating of the mass-loading, but soon reaches a peak
  temperature and at greater radii becomes cooler as the continued
  mass-loading shares out the thermal energy amongst more
  particles. Almost all of the gas in the bubble is substantially
  denser than the ambient intercloud gas, and is cold with
  temperatures of $\approx 20$\,K. High pressures arise during the
  initial frictional heating which we believe are ultimately
  responsible for the slightly higher momentum at late times.}.  Such
strong mass-loading is unlikely to occur in real WBBs.

To obtain the similarity solutions the mass injection rate from the
clumps must be radially dependent. For a stellar wind with a time
independent mechanical luminosity and an intercloud ambient medium of
constant density, it is required that $\dot{\rho} \propto
r^{-5/3}$. As $\dot{\rho}$ is not spatially dependent in our current
work (though it does decrease with time), further comparison to these
papers is unfortunately not possible.

\citet{Arthur:1993,Arthur:1996} used hydrodynamical simulations to
investigate mass-loading in the RCW\,58 WBB. They assumed that the
volumetric mass injection rate only depended on the Mach number of the
flow, and adopted the prescription of \citet{Hartquist:1986}. Since
the mass loading results in a fairly constant $M\approx0.6-0.7$ in the
bubble interior, it is clear that as the bubble grows, the global mass
injection rate increases in their simulations. This again differs from
our work, where the global rate of mass injection remains constant
(unless the clump mass reservoir runs out). It is unclear which of
these different scenarios best represents reality, and in any case the
specific clump distribution may vary from object to
object. Nevertheless, the same general effects due to the mass-loading
are observed.

\citet{Arthur:2012} simulated the Orion nebula as a combined WBB and
\hii region, including mass-loading from the embedded proplyds and
from thermal conduction at the edge of the hot bubble. The
mass-loading rate due to the proplyds was assumed to be radially
dependent, following the observed spatial density distribution of the
proplyds. A mixing efficiency of 10 per cent was assumed for the
injected mass, but since the post-shock bubble temperature was found
to be higher than observed, a higher mixing efficiency might be
appropriate. Alternatively, oblique shocks and/or the downstream
turbulence generated by many mass-loading sources on a flow may play a
role in reducing the post-shock temperature
\citep[][]{Pittard:2005,Aluzas:2012}.

In comparison to our new work, previous works which modelled
mass-loaded bubbles by treating clumps in the continuous limit have
two main shortcomings. Firstly, they allow for an infinite reservoir
of clump mass, which allows mass-injection to occur at all radii,
whereas in reality the bubbles are likely to become devoid of clumps
in their central regions as the clumps are destroyed (see, e.g.,
Fig.~3 in \citet{Rogers:2013} and Figs.~\ref{fig:wbb_finitereservoir}
and~\ref{fig:clump_survival}). Secondly, the Mach-number dependent
mass-injection rate used in \citet{Arthur:1993,Arthur:1996} is based
on an incorrect scaling \citep[see][]{Pittard:2010}.

We also note that the global normalization of the mass injection rate
is simply scaled in these earlier models, whereas in reality it is
likely to depend in some way on the stellar mass-loss rate or on the
stellar ionizing photon flux (if hydrodynamic ablation or
photoevaporation is the dominant mixing process, respectively). For
instance, Fig.~10 in \citet{Rogers:2013} shows that the mass-loading
factor of the outflow from a stellar cluster increases when each star
enters their Wolf-Rayet phase. Having said this, it is also clear that
the scaling is not necessarily a linear one and is also likely to be
time dependent.

\citet{McKee:1984} hypothesized that stellar wind bubbles are made
radiative by mass input from photoevaporating clumps, and for this
reason \citet{Matzner:2002} assumed that WBBs do not generate momentum
in excess of the wind momentum itself. In other words, mass-loading
``quenches'' the bubble. Recent numerical simulations of WBBs
expanding into a turbulent medium show that efficiently cooled bubbles
that approach momentum-conserving-like behaviour can exist
\citep{Lancaster:2021b}. However, examination of Fig.~10 in this work
reveals that in 7 out of 12 simulations the fractional turbulence
shows a significant drop before the bubble breaks out of the
simulation domain, while Fig.~8 shows that the momentum enhancement
factor in nearly all cases is rising with time. This perhaps opens the
door for a later transition to energy-conserving-like behaviour,
although in most cases it would likely not arise before the wind
bubble has broken out of its local cloud environment when the nature
of the bubble becomes drastically different. As noted in the
introduction, the cooling at the interface between hot and cold gas in
these simulations may also be overestimated.

\citet{Lancaster:2021a} find that due to efficient cooling in their
WBBs, the bubble pressure is substantially lower than that of the
standard \citet{Weaver:1977} bubble (see, e.g., their Fig.~2). In this
respect our results are quite different, as we find mass-loading with
$f_{\rm ML}=10^{2}$ slightly reduces the bubble pressure, but that
stronger mass-loading with $f_{\rm ML}=10^{3}$ first increases the
bubble pressure over the \citet{Weaver:1977} value, and then decreases
it (see panel g in
Figs.~\ref{fig:wbb_varfML}-\ref{fig:wbb_fML1e3_late}). Fig.~17 in
\citet{Lancaster:2021b} shows that the retained energy fraction in
their simulations is typically $\sim 0.1$ at early times, and
decreases with time, reaching $\sim 0.01$ at later times. This level
of cooling is stronger than in our low ambient density simulations,
where for $f_{\rm ML}=10^{3}$ this fraction is $0.2-0.3$ over most of
the bubble life (see Fig.~\ref{fig:wbb_varfML}h). Ultimately, this
difference allows our bubble to do more $PdV$ work and reach a
relatively higher radial momentum. However, in our simulations at
higher ambient density, strong mass-loading can cause the bubble to
radiate 98 per cent of the input energy. This finding is in better
agreement with those of \citet{Lancaster:2021b}, though again we find
a significant momentum boost.

Finally, we note that \citet{Pittard:2019} investigated mass-loading
in simulations of SNRs expanding into an inhomogeneous environment,
using the same approach as this work. Since the final radial momentum
was usually reduced by less than a factor of two, it appears that WBBs
are more sensitive to mass-loading than SNRs. On the other hand, we
note that heavily mass-loaded SNRs are not able to regenerate high
temperature gas if all the clumps within a specific region are
destroyed, unlike the behaviour we find for WBBs (see
Fig.~\ref{fig:wbb_finitereservoir}).

\subsection{Comparison to observations}
\label{sec:obs_comp}
The environment that bubbles expand into is typically clumpy, so they
are expected to undergo some form of mass-loading. Bubbles can be
blown by single massive stars which are either young or evolved, or by
groups of massive stars in stellar clusters. Let us first consider
young, single massive stars. One of the most interesting studies to
date is of N49, a dusty WBB blown by an O5V star with an age of
$0.5-1.0$\,Myr \citep{Everett:2010}. Because of the inferred short
lifetime of the dust, dusty gas is thought to be continuously injected
into the bubble by high density clouds
($n \sim 10^{5}\,{\rm cm^{-3}}$) that are overrun and engulfed. The
mass-loss rate of the central star is estimated as
$\Mdot_{\rm w} = 1.5\times10^{-6}\,\Msolpyr$, while the clump
injection rate is estimated as
$\Mdot_{\rm cl} = 2.5\times10^{-4}\,\Msolpyr$. This gives a
mass-loading strength
$f_{\rm ML} = \Mdot_{\rm cl}/\Mdot_{\rm w} \approx 150$. The total
mass injected by the clumps so far is thought to be
$125-250\,\Msol$. The external intercloud density is not well
constrained but estimated to have a number density
$n_{\rm H} \sim 10^{4}\,{\rm cm^{-3}}$. Since the radius of the bubble
is 2\,pc, the swept-up mass is $1.12\times10^{4}\,\Msol$. This gives a
minimum value for the clump to interclump mass ratio
$\nu = 125/11200 = 0.011$.  Comparison with Fig.~\ref{fig:higherramb}
suggests that the mass-loading in N49 might significantly affect the
bubble properties, though whether this is actually the case will
depend on whether the available reservoir of clump mass is large
enough (cf. Fig.~\ref{fig:wbb_finitereservoir}).

\citet{Everett:2010} also note that if the dust is gradually
evaporated there should be bubbles with a central dust-free region,
where the $24\,\mu$m emission from the injected dust is confined to a
bright rim.  It would be interesting to perform a full radiation
hydrodynamics model of the combined \hii region and WBB of N49, with
mass-loading from dusty embedded clumps, to compare to the available
data. However, this is beyond the scope of the current paper.

Indirect evidence for mass-loading in WBBs also comes from the lower
than expected X-ray temperatures that have been measured. However,
X-ray emission is currently very difficult to detect in WBBs produced
by single unevolved stars. No X-ray emission was detected from the
iconic Bubble Nebula (NGC\,7635), for instance
\citep{Toala:2020}. However, X-ray emission has been detected in the
Extended Orion Nebula, which is powered mainly by the star
$\theta^{1}$Ori\,C \citep{Gudel:2008}, and in the WBB around
$\zeta$\,Oph \citep{Toala:2016a}.

X-ray emission is more readily detected in WR nebulae, though only 4
have detected X-ray emission to date: S\,308, NGC\,2359, NGC\,3199,
and NGC\,6888, around WR\,6, WR\,7, WR\,18 and WR\,136, respectively
\citep[e.g.][]{Toala:2012,Toala:2014,Toala:2015,Toala:2016b,Toala:2017}. The
properties of the emission, including its relative softness,
brightness, and the inferred abundances and estimated electron density
of the hot gas, favour a scenario in which strong mixing of
circumstellar material from the outer shell (formed from the sweeping
up of a previous, slower wind) into the bubble interior occurs. This
process may be aided by the fragmentation of the shell, which results
in dense clumps becoming embedded in the hot interior gas
\citep[e.g.][]{Toala:2011}. Such a scenario is not radically different
from the work presented here: the main difference is that the clumps
originate from previous mass-loss from the star rather than from the
wider interstellar medium.

Finally, we note that X-ray emission has also been detected in young
(pre-SN) massive clusters, such as M17 and the Rosette Nebula
\citep{Townsley:2003}. It is thought that the X-ray emission arises
from the collective thermalization of the stellar winds, and softened
by mass-loading from embedded clumps and adjacent colder surfaces
\citep{Townsley:2011a}. In some objects the nearest embedded clumps
may have been destroyed/cleared away, with ongoing mass-loading of the
flow occuring only at greater distances.  Such faint diffuse X-ray
emission seems to be a ubiquitous property of massive star forming
regions
\citep[e.g.][]{Townsley:2011b,Townsley:2014,Townsley:2018,Townsley:2019}.
Dedicated modelling of specific clusters is needed to make further
progress, such as has been attempted for M17
\citep{Reyes-Iturbide:2009,Velazquez:2013} and the Rosette Nebula
\citep{Wareing:2018}.

\section{Summary and conclusions}
\label{sec:summary}
We have examined the properties and behaviour of wind-blown bubbles
expanding into a clumpy, inhomogeneous medium. The expanding bubble is
assumed to sweep up intercloud material, and to sweep over
pre-existing clouds which are destroyed within it as they become
overrun/engulfed. The cloud destruction adds mass into the bubble,
which we assume rapidly merges with the global flow and attains the
same density, velocity and temperature. We assume that the mixing
timescale of the gas is much shorter than the cooling timescale of the
mixing gas, so that there is no significant cooling during this
transition.  The nature of the mass-loading is parameterized by two
variables: the mass-loading strength,
$f_{\rm ML} = \Mdot_{\rm cl}/\Mdot_{\rm w}$, and the ratio of cloud to
intercloud mass per unit volume in the ambient medium, $\nu$. The mass
injection is assumed to occur uniformly within the bubble, unless and
until the available mass reservoir at a particular radius is
exhausted.

We find that:
\begin{enumerate}
\item Mass injection can affect the behaviour and evolution of the
  bubble from its earliest stages. It increases the density and
  decreases the velocity within the bubble. In the pre-shock stellar
  wind it increases the temperature through drag heating, while it
  reduces the temperature of hot shocked gas as the available energy
  is shared between more particles. The affect of mass-loading on the
  volume-averaged pressure in the bubble is more complicated, and may
  increase or decrease it (in some cases this depends also on the
  bubble age). However, mass-loading always enhances the radiative
  cooling and reduces the retained energy fraction.
\item Mass-loaded bubbles do not expand as quickly or as far. They
  cool more quickly, do less $PdV$ work on the swept-up gas, and
  ultimately attain a lower final momentum. However, they can still
  provide a significant boost to the radial momentum input by the
  wind. This is especially true if the mass-loading is relatively weak
  and/or the available mass in clouds is relatively low. However, even
  when cooling losses become severe, and the retained energy fraction
  drops to very low values, we find that the bubble may still
  substanially boost the wind momentum. In this respect, our
  mass-loaded bubbles behave more like energy-conserving bubbles,
  rather than the momentum-conserving-like behaviour of ``quenched''
  bubbles.
\item If the available clump mass is limited and starts to run out,
  the reduction to the final radial momentum is not as severe. In some
  cases, parts of the bubble may become clump free and not subject to
  any current mass-loading, while other parts may still contain clumps
  and continue to be mass-loaded. This can create interesting density
  and temperature profiles. In such cases, high temperature gas can be
  regenerated (unlike in SNRs).
\item Mass-loading also drag heats the stellar wind prior to its
  thermalization at the reverse shock. In extreme cases the reverse
  shock no longer exists, though this is unlikely to occur in real
  bubbles.
\end{enumerate} 

In summary, mass-loading can significantly affect the behaviour of
WBBs. However, we find that for the model parameters explored in this
work, the bubbles can still perform significant $PdV$ work on the
surrounding gas, and provide substantial boosts to the radial momentum
input by the wind.

\section*{Acknowledgements}
We thank Joshua Selby for running some simulations in the very early
stages of this work, and the referee for very detailed and useful
suggestions.  We acknowledge support from the Science and Technology
Facilities Council (STFC, Research Grant ST/P00041X/1).

\section*{Data Availability}
The data underlying this article are available in the Research
Data Leeds Repository, at \url{https://doi.org/10.5518/1183}.








\bsp	
\label{lastpage}
\end{document}